\documentstyle[twoside]{article}
\catcode`\@=11
\long\def\@makefntext#1{
\protect\noindent \hbox to 3.2pt {\hskip-.9pt
$^{{\eightrm\@thefnmark}}$\hfil}#1\hfill}               

\def\@makefnmark{\hbox to 0pt{$^{\@thefnmark}$\hss}}    

\def\ps@myheadings{\let\@mkboth\@gobbletwo
\def\@oddhead{\hbox{}
\rightmark\hfil\eightrm\thepage}
\def\@oddfoot{}\def\@evenhead{\eightrm\thepage\hfil
\leftmark\hbox{}}\def\@evenfoot{}
\def\sectionmark##1{}\def\subsectionmark##1{}}

\oddsidemargin=\evensidemargin
\newcounter{sectionc}\newcounter{subsectionc}\newcounter{subsubsectionc}
\renewcommand{\section}[1] {\vspace{12pt}\addtocounter{sectionc}{1}
\setcounter{subsectionc}{0}\setcounter{subsubsectionc}{0}\noindent
        {\tenbf\thesectionc. #1}\par\vspace{5pt}}
\renewcommand{\subsection}[1] {\vspace{12pt}\addtocounter{subsectionc}{1}
        \setcounter{subsubsectionc}{0}\noindent
        {\bf\thesectionc.\thesubsectionc. {\kern1pt \bfit #1}}\par\vspace{5pt}}
\renewcommand{\subsubsection}[1] {\vspace{12pt}\addtocounter{subsubsectionc}{1}
        \noindent{\tenrm\thesectionc.\thesubsectionc.\thesubsubsectionc.
        {\kern1pt \tenit #1}}\par\vspace{5pt}}
\newcommand{\nonumsection}[1] {\vspace{12pt}\noindent{\tenbf #1}
        \par\vspace{5pt}}

\newcounter{appendixc}
\newcounter{subappendixc}[appendixc]
\newcounter{subsubappendixc}[subappendixc]
\renewcommand{\thesubappendixc}{\Alph{appendixc}.\arabic{subappendixc}}
\renewcommand{\thesubsubappendixc}
        {\Alph{appendixc}.\arabic{subappendixc}.\arabic{subsubappendixc}}

\renewcommand{\appendix}[1] {\vspace{12pt}
        \refstepcounter{appendixc}
        \setcounter{figure}{0}
        \setcounter{table}{0}
        \setcounter{lemma}{0}
        \setcounter{theorem}{0}
        \setcounter{corollary}{0}
        \setcounter{definition}{0}
        \setcounter{equation}{0}
        \renewcommand{\thefigure}{\Alph{appendixc}.\arabic{figure}}
        \renewcommand{\thetable}{\Alph{appendixc}.\arabic{table}}
        \renewcommand{\theappendixc}{\Alph{appendixc}}
        \renewcommand{\thelemma}{\Alph{appendixc}.\arabic{lemma}}
        \renewcommand{\thetheorem}{\Alph{appendixc}.\arabic{theorem}}
        \renewcommand{\thedefinition}{\Alph{appendixc}.\arabic{definition}}
        \renewcommand{\thecorollary}{\Alph{appendixc}.\arabic{corollary}}
        \renewcommand{\theequation}{\Alph{appendixc}.\arabic{equation}}
        \noindent{\tenbf Appendix \theappendixc #1}\par\vspace{5pt}}
\newcommand{\subappendix}[1] {\vspace{12pt}
        \refstepcounter{subappendixc}
        \noindent{\bf Appendix \thesubappendixc. {\kern1pt \bfit #1}}
        \par\vspace{5pt}}
\newcommand{\subsubappendix}[1] {\vspace{12pt}
        \refstepcounter{subsubappendixc}
        \noindent{\rm Appendix \thesubsubappendixc. {\kern1pt \tenit #1}}
        \par\vspace{5pt}}

\newcommand{\textlineskip}{\baselineskip=13pt}
\newcommand{\smalllineskip}{\baselineskip=10pt}

\def\eightcirc{
\begin{picture}(0,0)
\put(4.4,1.8){\circle{6.5}}
\end{picture}}
\def\eightcopyright{\eightcirc\kern2.7pt\hbox{\eightrm c}}

\newcommand{\copyrightheading}[1]
        {\vspace*{-2.5cm}\smalllineskip{\flushleft
{\bf OUT--4102--54} \hfill          
{\bf hep-ph/9504398}
\hfill{\bf INP--95--13/377}         
\\}}

\def\abstracts#1#2#3{{
        \centering{\begin{minipage}{4.5in}\baselineskip=10pt\footnotesize
        \parindent=0pt #1\par
        \parindent=15pt #2\par
        \parindent=15pt #3
        \end{minipage}}\par}}

\newcommand{\bibit}{\nineit}
\newcommand{\bibbf}{\ninebf}
\renewenvironment{thebibliography}[1]
        {\frenchspacing
         \ninerm\baselineskip=11pt
         \begin{list}{\arabic{enumi}.}
        {\usecounter{enumi}\setlength{\parsep}{0pt}
         \setlength{\leftmargin 17pt}{\rightmargin 0pt}   
         \setlength{\itemsep}{0pt} \settowidth
        {\labelwidth}{#1.}\sloppy}}{\end{list}}

\newcounter{itemlistc}
\newcounter{romanlistc}
\newcounter{alphlistc}
\newcounter{arabiclistc}

\newcommand{\fcaption}[1]{
        \refstepcounter{figure}
        \setbox\@tempboxa = \hbox{\footnotesize Fig.~\thefigure. #1}
        \ifdim \wd\@tempboxa > 5in
           {\begin{center}
        \parbox{5in}{\footnotesize\smalllineskip Fig.~\thefigure. #1}
            \end{center}}
        \else
             {\begin{center}
             {\footnotesize Fig.~\thefigure. #1}
              \end{center}}
        \fi}

\newcommand{\tcaption}[1]{
        \refstepcounter{table}
        \setbox\@tempboxa = \hbox{\footnotesize Table~\thetable. #1}
        \ifdim \wd\@tempboxa > 5in
           {\begin{center}
        \parbox{5in}{\footnotesize\smalllineskip Table~\thetable. #1}
            \end{center}}
        \else
             {\begin{center}
             {\footnotesize Table~\thetable. #1}
              \end{center}}
        \fi}

\def\@citex[#1]#2{\if@filesw\immediate\write\@auxout
        {\string\citation{#2}}\fi
\def\@citea{}\@cite{\@for\@citeb:=#2\do
        {\@citea\def\@citea{,}\@ifundefined
        {b@\@citeb}{{\bf ?}\@warning
        {Citation `\@citeb' on page \thepage \space undefined}}
        {\csname b@\@citeb\endcsname}}}{#1}}

\newif\if@cghi
\def\cite{\@cghitrue\@ifnextchar [{\@tempswatrue
        \@citex}{\@tempswafalse\@citex[]}}
\def\citelow{\@cghifalse\@ifnextchar [{\@tempswatrue
        \@citex}{\@tempswafalse\@citex[]}}
\def\@cite#1#2{{$\null^{#1}$\if@tempswa\typeout
        {IJCGA warning: optional citation argument
        ignored: `#2'} \fi}}

\def\pmb#1{\setbox0=\hbox{#1}
        \kern-.025em\copy0\kern-\wd0
        \kern.05em\copy0\kern-\wd0
        \kern-.025em\raise.0433em\box0}

\def\fnt#1#2{\footnotetext{\kern-.3em
        {$^{\mbox{\scriptsize #1}}$}{#2}}}

\def\fpage#1{\begingroup
\voffset=.3in
\thispagestyle{empty}\begin{table}[b]\centerline{\footnotesize #1}
        \end{table}\endgroup}

\def\runninghead#1#2{\pagestyle{myheadings}
\markboth{{\protect\footnotesize\it{\quad #1}}\hfill}
{\hfill{\protect\footnotesize\it{#2\quad}}}}
\font\tenrm=cmr10
\font\tenit=cmti10
\font\tenbf=cmbx10
\font\bfit=cmbxti10 at 10pt
\font\ninerm=cmr9
\font\nineit=cmti9
\font\ninebf=cmbx9
\font\eightrm=cmr8

\def\qed{\hbox{${\vcenter{\vbox{                        
   \hrule height 0.4pt\hbox{\vrule width 0.4pt height 6pt
   \kern5pt\vrule width 0.4pt}\hrule height 0.4pt}}}$}}

\def\bsc{{\sc a\kern-6.4pt\sc a\kern-6.4pt\sc a}}       
\def\bflatex{\bf L\kern-.30em\raise.3ex\hbox{\bsc}\kern-.14em
T\kern-.1667em\lower.7ex\hbox{E}\kern-.125em X}
\begin{document}
\runninghead
{Three--loop QED Vacuum Polarization  \ldots}
{\ldots\ Four--loop Muon Anomalous Magnetic Moment}
\normalsize\textlineskip
\thispagestyle{empty}
\setcounter{page}{1}
\copyrightheading{}                     
\vspace*{0.88truein}
\fpage{1}
\centerline{\large\bf
Three--loop QED Vacuum Polarization and the}
\vspace*{0.035truein}
\centerline{\large\bf
Four--loop Muon Anomalous Magnetic Moment\footnote{
Presented by P.\ A.\ Baikov,
at the AI--HENP 95 workshop, Pisa, April 1995}
\footnote{INP--OU collaboration, supported in part by INTAS project
93--1180 (contract 1010--CT93--0024)}}
\vspace*{0.37truein}
\centerline{\footnotesize
P.\ A.\ BAIKOV\footnote{
Supported in part
by the Russian Basic Research Foundation (grant N 93--02--14428);\\
\hspace*{3pt}Email: baikov@theory.npi.msu.su}}
\vspace*{0.015truein}
\centerline{\footnotesize\it Institute of Nuclear Physics,
Moscow State University}
\baselineskip=10pt
\centerline{\footnotesize\it  119~899, Moscow, Russia}
\vspace*{10pt}
\centerline{\normalsize and}
\vspace*{10pt}
\centerline{\footnotesize
D.~J.~BROADHURST\footnote{
Email: D.Broadhurst@open.ac.uk}}
\vspace*{0.015truein}
\centerline{\footnotesize\it
Physics Department, Open University}
\baselineskip=10pt
\centerline{\footnotesize\it
Milton Keynes, MK7 6AA, UK}
\vspace*{0.225truein}
\vspace*{0.21truein}
\abstracts{
Three--loop contributions to massive QED vacuum polarization are
evaluated by a combination of analytical and numerical techniques.
The first three Taylor coefficients, at small $q^2$, are obtained
analytically, using $d$\/--dimensional recurrence relations.
Combining these with analytical input at threshold, and at
large $q^2$, an accurate Pad\'e approximation is obtained, for all $q^2$.
Inserting this in the one--loop diagram for the muon anomalous
magnetic moment, we find reasonable agreement with four--loop,
single--electron--loop, muon--anomaly contributions,
recently re--evaluated by Kinoshita,
using 8--dimensional Monte--Carlo integration.
We believe that our new method is at least two orders of magnitude
more accurate than the Monte--Carlo approach,
whose uncertainties appear to have been underestimated,
by a factor of 6.
}{}{}
\vspace*{1pt}\textlineskip      
\section{Introduction}          
\vspace*{-0.5pt}\noindent
We describe a method,
previously tested\cite{GG2} in two--loop QCD,
to approximate, to high accuracy,
three--loop contributions to QED vacuum polarization,
using new analytical results for the small momentum--transfer limit,
combined with asymptotic\cite{GKL1,GKL2,GKL3} and threshold\cite{SV}
results.
Related contributions to the four--loop muon anomalous magnetic
moment are computed, to test an evaluation\cite{TK} that was undertaken
in response to a previous discrepancy between numerical\cite{KNO}
and analytical\cite{BKT} work.

In the on--shell (OS) renormalization scheme of conventional QED,
the renormalized photon propagator has a denominator ($1+\Pi(z)$),
where $z\equiv q^2/4m^2$, with an electron mass $m$, and
the vacuum polarization function, $\Pi(z)$, vanishes at $z=0$.
Non--relativistic consideration of the electron--positron system
yields information about the threshold\cite{SV} behaviour, as $z\to1$.
Moreover, the $\overline{\rm MS}$ asymptotic behaviour\cite{GKL2,GKL3},
as $z\to-\infty$, combined with relations\cite{BKT,REC} between the
$\overline{\rm MS}$ and OS schemes, yields two terms of the asymptotic
expansion in powers of $1/z$.

The crucial new ingredient is our use of recurrence relations\cite{REC},
to obtain the first three terms of the expansion as $z\to0$.
Combining 6 analytical data with Pad\'e\cite{GG2,BFT}
(or hypergeometric\cite{GG2,BIS}) approximations, we shall
produce reliable fits, for all $z$,
and hence check four--loop muon--anomaly contributions\cite{TK}.

\section{Small--momentum behaviour}
\vspace*{-0.5pt}\noindent
We evaluated, to 3 loops, the first 3 moments in the $z\to 0$ expansion
\[\Pi(z)  =  \sum_{n>0} C_n\,z^n + {\rm O}(\alpha^4)\,,\]
by intensive application of $d$\/--dimensional recurrence
relations to three--loop massive vacuum
diagrams\cite{REC}, with propagators raised to powers up to 11,
since up to 8 differentiations w.r.t.\ the external momentum $q$
are required before setting it to zero. This put great demands
on the REDUCE package RECURSOR\cite{REC}, which used 80~MB
of memory, for 2 days, on a DecAlpha machine, after hand--tuning the
procedures, to minimize recomputation of integrals, and to allow safe
truncation in $\varepsilon=(4-d)/2$. The gauge invariance
of $C_1$ and $C_2$ was verified for all $\varepsilon$. After OS
mass\cite{GBGS,BGS} and charge\cite{REC} renormalization,
we obtained the finite $\varepsilon\to0$ limits
\newcommand{\Df}[2]{\mbox{$\frac{#1}{#2}$}}
\begin{eqnarray}
C_1 & = & \Bigl\{ N^2\left[ \Df{8}{15}                            \,\zeta_2
                          + \Df{203}{864}                         \,\zeta_3
                          - \Df{11407}{11664}       \right]\nonumber\\
    &   &{}      +N  \left[             \left(1-\Df{8}{5}\ln2\right)\zeta_2
                          + \Df{22781}{6912}                      \,\zeta_3
                          - \Df{8687}{3456}                  \right]
                           \Bigr\}\frac{\alpha^3}{\pi^3}  \label{C1}\\
    &   &{}+   \Df{82}{81}     \,N\frac{\alpha^2}{\pi^2}
           +   \Df{4}{15}      \,N\frac{\alpha  }{\pi  }\,,\nonumber\\
C_2 & = & \Bigl\{ N^2\left[ \Df{16}{35}                           \,\zeta_2
                          + \Df{14203}{73728}                     \,\zeta_3
                          - \Df{1520789}{1658880}   \right]\nonumber\\
    &   &{}      +N  \left[ \Df{6}{7}   \left(1-\Df{8}{5}\ln2\right)\zeta_2
                          + \Df{4857587}{184320}                  \,\zeta_3
                          - \Df{223404289}{7464960}          \right]
                           \Bigr\}\frac{\alpha^3}{\pi^3}  \label{C2}\\
    &   &{}+   \Df{449}{675}   \,N\frac{\alpha^2}{\pi^2}
           +   \Df{4}{35}      \,N\frac{\alpha  }{\pi  }\,,\nonumber\\
C_3 & = & \Bigl\{ N^2\left[ \Df{128}{315}                         \,\zeta_2
                          + \Df{12355}{55296}                     \,\zeta_3
                          - \Df{83936527}{93312000})\right]\nonumber\\
     &  &{}     + N  \left[ \Df{16}{21} \left(1-\Df{8}{5}\ln2\right)\zeta_2
                          + \Df{33067024499}{206438400}           \,\zeta_3
                          - \Df{885937890461}{4644864000}    \right]
                           \Bigr\}\frac{\alpha^3}{\pi^3}  \label{C3}\\
     &  &{}+\Df{249916}{496125}\,N\frac{\alpha^2}{\pi^2}
           +\Df{64}{945}       \,N\frac{\alpha  }{\pi  }\,,\nonumber
\end{eqnarray}
where we follow common practice\cite{BKT}, by allowing for
$N$ degenerate leptons. In pure QED, $N=1$; formally,
the powers of $N$ serve to count the number of electron loops.
Our principal interest, for consideration of four--loop muon--anomaly
contributions\cite{TK},
is $\Pi_3^{[1]}\alpha^3/\pi^3$, the three--loop contributions
to $\Pi$ that involve a {\it single\/} electron loop. The moments
of $\Pi_3^{[1]}(z)$ are given
by the coefficients of $N\alpha^3/\pi^3$. (We shall not
need the $N^2\alpha^3/\pi^3$ terms in Section~6, since the muon--anomaly
contributions of the diagrams with two electron loops are
better understood\cite{BKT,LNF}.)

\newpage
\section{Large--momentum behaviour}
\vspace*{-0.5pt}\noindent
The situation regarding the $z\to-\infty$ behaviour of $\Pi(z)$
was unclear, until recently, because
three--loop $\overline{\rm MS}$ QCD results\cite{GKL1} had been
altered, while obtaining QED results\cite{GKL2}, in the belief
(now known\cite{GKL3} to be mistaken) that the former contained errors.
Further calculation\cite{KGC} confirmed the QCD results\cite{GKL1}
and hence invalidated the ${\rm O}(1/z)$ QED results\cite{GKL2}.
Accordingly, we thought it prudent to derive the OS asymptotic
behaviour ourselves, from first principles, using the
REDUCE package SLICER, which had been written specifically to
check\cite{BKT} the leading, massless, $\overline{\rm MS}$ behaviour,
obtained\cite{GKL2} with the SCHOONSCHIP package
MINCER\cite{GLST}.

In our {\it ab initio\/} derivation of the asymptotic OS result for
$\Pi_3^{[1]}$, we used neither the $\overline{\rm MS}$ scheme, nor MINCER.
Instead, the asymptotic expansion of the bare diagrams was obtained,
in $d$ dimensions, using SLICER, and the bare charge and mass were transformed
directly to the physical charge and mass, using multiplicative OS
renormalizations\cite{REC,BGS}, obtained by RECURSOR.
Setting $\varepsilon=0$, we obtained a finite OS result of the form
\[\Pi_3^{[1]}(z)=A(z)+B(z)/z+{\rm O}(L^3/z^2)\,,\]
where $L\equiv\ln(-4z)=\ln(-q^2/m^2)$ and
\begin{eqnarray}
A(z)&=&
-\Df{121}{192}
+\Df{5}{2}\zeta_5
-\Df{99}{64}\zeta_3
+2\left(\ln2-\Df{5}{8}\right)\zeta_2
+\Df{1}{32}L\,,
\label{A}\\
B(z)&=&\,\,
\Df{139}{48}
-\Df{35}{24}\zeta_5
-\Df{41}{48}\zeta_3
+3\left(\ln2-\Df{5}{8}\right)\zeta_2
-\Df{3}{32}\left(L-6L^2\right)\,.
\label{B}
\end{eqnarray}
Using finite transformations\cite{REC,GBGS} from physical to
$\overline{\rm MS}$--renormalized quantities, one obtains, from our
OS result, an $\overline{\rm MS}$ asymptotic behaviour identical
to that which would\cite{ALK} have been obtained from the QCD
analysis\cite{GKL1}, had it not been miscorrected in the course
of deriving QED results\cite{GKL2}. As a result
of our, and other\cite{KGC}, work, a (second) erratum\cite{GKL3}
to the QED work\cite{GKL2} was issued.

In conclusion, we are
confident of our OS QED result, since it is quite independent
of previous works\cite{GKL1,GKL2,KGC} and, eventually\cite{GKL3,ALK},
in agreement with them.

\section{Threshold behaviour}
\vspace*{-0.5pt}\noindent
The leading threshold behaviour, at 3 loops,
is determined by non--relativistic quantum mechanics:
$\Pi_3^{[1]}(z)=\frac{1}{24}\pi^5(1-z)^{-1/2}+{\rm O}(\ln(1-z))$, as $z\to1$.
Moreover, it appears\cite{SV,MBV} that a stronger statement can be made,
namely that the first relativistic correction to the spectral
density, $\rho(t)\equiv{\rm Im}\,\Pi(t+{\rm i}0)/\pi$, at any given order
in $\alpha$, is cancelled in the combination $(1+4\alpha/\pi)\rho(t)$.
At the two--loop level, the exact relativistic
results confirm that $\rho_2(t)+4\,\rho_1(t)=\pi^2+{\rm O}(v^2)$ is free
of a term of first order in $v\equiv(1-1/t)^{1/2}$. The corresponding
cancellation at 3 loops is expected\cite{MBV} to occur in
$v(\rho_3^{[1]}+4\,\rho_2)=\frac{1}{24}\pi^4+{\rm O}(v^2)$,
implying that
\begin{equation}
\lim_{z\to1}\left(\Pi_3^{[1]}(z)+4\,\Pi_2(z)
-\frac{\pi^5}{24(1-z)^{1/2}}\right)=\mbox{constant}\,,
\label{C}
\end{equation}
with an unknown value for the constant, but no logarithmic singularity.

\newpage
\newcommand{\Pit}{\tilde{\Pi}_3^{[1]}}
\section{Approximation method}
\vspace*{-0.5pt}\noindent
We express the analytical results~(\ref{C1}--\ref{C})
as properties of the combination
\begin{equation}
\Pit(z)\equiv\Pi^{[1]}_3(z)+4\,\Pi_2(z)
+(1-z)\,G(z)\left(\frac{9}{4}\,G(z)+\frac{31}{16}+\frac{229}{32z}\right)
-\frac{229}{32z}-\frac{173}{96}\,,
\label{P}
\end{equation}
where $G(z)\equiv{}_2F_1\left(1,1;\frac32;z\right)$ is given by
$(z^2-z)^{-1/2}{\rm arcsinh}(-z)^{1/2}$,
on the negative real axis,
and the two--loop term, $\Pi_2(z)$, is quadratic in
$G(z)$ and involves a trilogarithm\cite{BKT} and its derivative.

The data are conveniently encoded by the moments
\[M(n)\equiv\int_1^\infty\frac{{\rm d}t}{t^{n+1}}\,\tilde\rho_3^{[1]}(t)\]
of the spectral density of $\Pit$. At small $z$, we have $\Pit(z)=
\sum_{n>0}M(n)\,z^n$ and hence obtain $M(1)$, $M(2)$, $M(3)$
from the coefficients of $N\alpha^3/\pi^3$ in the results
of Eqs~(\ref{C1},\ref{C2},\ref{C3}) for the moments of $\Pi(z)$,
after taking account of the known moments of the
additional terms in Eq~(\ref{P}). At large $z$,
the logarithmic singularities of these additional terms cancel,
by deliberate construction, those of Eqs~(\ref{A},\ref{B}), whose constant
terms therefore determine $M(0)$ and $M(-1)$, respectively.
Finally, the threshold Coulomb singularity of Eq~(\ref{C})
gives the large--$n$ behaviour of the ratio
\[R(n)\equiv\frac{M(n)}{C(n)}=\frac{\pi^4}{24}
+{\rm O}(1/n)\,,
\qquad C(n)\equiv\int_1^{\infty}\frac{{\rm d}t}{t^{n+1}}
\,\rho_3^{[{\rm c}]}(t)
=\frac{(1)_{n+1}}{\left(\frac12\right)_{n+2}}\,,\]
where $C(n)$ is the moment of a spectral density
$\rho_3^{[{\rm c}]}(t)\equiv t^{-3/2}(t-1)^{-1/2}$,
with a coulombic $1/v$ threshold singularity and the same convergence
properties, at large $t$, as $\tilde{\rho}_3^{[1]}(t)$.
Note that a further datum, namely
the absence of a logarithmic singularity in Eq~(\ref{C}),
corresponds to the absence of an ${\rm O}(1/n^{1/2})$ term in $R(n)$,
as $n\to\infty$, partly accounting for the remarkable uniformity
of our final analytical database:
\[\begin{array}{rll}
R(-1)  = &
-\frac{3}{2}\left(\ln2-\frac{5}{8}\right)\zeta_2+
\frac{41}{96}\zeta_3+\frac{35}{48}\zeta_5+\frac{911}{384} &{}=
3.473\,721\,028\,889 \\[3pt]
R(0)  = &
-\frac{3}{2}\left(\ln2-\frac{5}{8}\right)\zeta_2+
\frac{1065}{256}\zeta_3-\frac{15}{8}\zeta_5+\frac{307}{256} &{}=
4.087\,577\,319\,347 \\[3pt]
R(1)  = &
-\frac{3}{2}\left(\ln2-\frac{5}{8}\right)\zeta_2+
\frac{113905}{36864}\zeta_3+\frac{35551}{55296} &{}=
4.188\,975\,919\,282 \\[3pt]
R(2)  = &
-\frac{3}{2}\left(\ln2-\frac{5}{8}\right)\zeta_2+
\frac{34003109}{1179648}\zeta_3-\frac{7229411363}{238878720} &{}=
4.216\,954\,083\,059 \\[3pt]
R(3)  = &
-\frac{3}{2}\left(\ln2-\frac{5}{8}\right)\zeta_2+
\frac{33067024499}{167772160}\zeta_3-\frac{6144308789323}{26424115200} &{}=
4.224\,481\,581\,719 \\[3pt]
R(\infty)  = &
\frac{15}{4}\zeta_4 &{}=
4.058\,712\,126\,417
\end{array}\]
which has been obtained from 3 quite disparate regimes.

Now we map\cite{GG2} the cut $z$\/--plane onto the unit disk,
and define a mapped function
\begin{equation}
P(\omega)\equiv\frac{1-\omega}{(1+\omega)^2}
\left(\Pit(z)-\Pit(\infty)\right)\,,
\qquad z=\frac{4\omega}{(1+\omega)^2}\,,
\label{M}
\end{equation}
which is analytic for $|\omega|<1$, with the cut mapped to the
unit circle.
The 6 data then determine
$\{P(-1),P(0),
P^\prime(0),
P^{\prime\prime}(0),
P^{\prime\prime\prime}(0),
P(1)\}$, allowing us to construct [2/3] and [3/2] Pad\'e
approximants\cite{GG2}, with benign poles outside the unit disk,
and imaginary parts on the unit circle that accurately approximate
the spectral density.
The differences between these two approximations are very small,
for all $|\omega|\leq1$.

\newpage
\newcommand{\am}{A_4^{[1]}}
\section{Four--loop contribution to the Muon Anomalous Magnetic Moment}
\vspace*{-0.5pt}\noindent
Our simple rational approximations to $P(\omega)$ reproduce, exactly,
all known data on $\Pi_3^{[1]}$, as well as its
analyticity structure. We now use them to calculate the
four--loop contribution\cite{TK}, $a_\mu=\am\alpha^4/\pi^4$, to
the muon anomaly, $(g/2-1)_\mu$, due to insertion of three--loop,
single--electron--loop vacuum polarization diagrams into the
one--loop anomaly diagram. A typical diagram is
\vspace{-7mm}
\setlength{\unitlength}{0.0141cm}
\newbox\shell

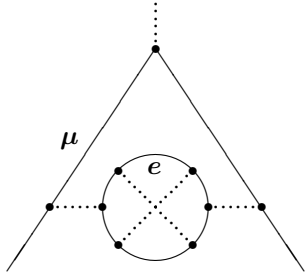
\def\pixel{\circle*{3}}
\def\vertx{\circle*{7}}
\def\gpict{\begin{picture}(400,390)(-100,-125)
           \put(0,0){\circle{100}}
           \put(-90,60){\boldmath$\mu$}
           \put(0,150){\line( 2,-3){140}}
           \put(0,150){\line(-2,-3){140}}
           \multiput(-50,0)(-7,0){8}\pixel
           \multiput( 50,0)( 7,0){8}\pixel
           \multiput(0,150)( 0,7){7}\pixel
           \put(   0,150){\vertx}
           \put( -50,  0){\vertx}
           \put(  50,  0){\vertx}
           \put( 100,  0){\vertx}
           \put(-100,  0){\vertx}}
\def\spict{\gpict
           \put( -8,32){\boldmath$e$}}
\def\epict{\end{picture}}
\def\ntype{\multiput(0,0)( 4.9, 4.9){8}\pixel
           \multiput(0,0)( 4.9,-4.9){8}\pixel
           \multiput(0,0)(-4.9, 4.9){8}\pixel
           \multiput(0,0)(-4.9,-4.9){8}\pixel
           \put( 35, 35){\vertx}
           \put(-35, 35){\vertx}
           \put( 35,-35){\vertx}
           \put(-35,-35){\vertx}}
\def\align{\dimen0=\ht\shell
           \multiply\dimen0by7
   \divide\dimen0by16
   \raise-\dimen0\box\shell}

\hspace*{25ex}
\setbox\shell=\hbox{\spict\ntype
\epict}\align

\vspace{-7mm}
\noindent
The resulting coefficient of $\alpha^4/\pi^4$ in the muon anomaly
is given by\cite{LNF}
\begin{equation}
\am=-\int_0^1 {\rm dy}\,(1-y)\,\Pi^{[1]}_3(z)\,,\qquad
z=-\,\frac{m_{\mu}^2}{4m^2}\,\frac{y^2}{1-y}\,.
\label{I}
\end{equation}
We calculate the integral using [3/2], [2/3], and [2/2] Pad\'e approximants
to $P(\omega)$. In the [2/2] approximants we omit a piece of data
from each regime, obtaining
\[\begin{array}{l|c|c|c|c|c|}
\mbox{Input}
&\mbox{all}
&\mbox{all}
&\mbox{omit Eq~(\ref{C3})}
&\mbox{omit Eq~(\ref{B})}
&\mbox{omit Eq~(\ref{C})}
\\[3pt]
\mbox{Pad\'e}
&[3/2]
&[2/3]
&[2,2]
&[2/2]
&[2/2]\\
-\am
&0.230\,362\,20
&0.230\,362\,18
&0.230\,360\,42
&0.230\,366\,94
&0.230\,361\,49
\end{array}\]
with a muon mass $m_{\mu}=206.768\,262\,m$. The stability is remarkable:
changing the Pad\'e method from [3/2] to [2/3] changes the output
by 1~part in $10^7$; removing a piece of data, from any of the 3
regimes, changes it by no more than 2~parts in $10^5$. The improvement
from using 6 inputs, as opposed to 5,
is greatest in the case
of including the asymptotic result of Eq~(\ref{B}).
In contrast, the Coulomb datum, $R(\infty)
=\frac{1}{24}\pi^4$, improves the result by only 3~parts in $10^6$,
since the muon--anomaly integral~(\ref{I}) involves only space--like
momenta. The smallness of our spread of results demonstrates
a high degree of consistency in the input, making the possibility of
analytical error very remote. Being conservative, we
take the range of [2/2] results as a measure of our uncertainty,
and arrive at $\am=-0.230362(5)$, to be compared
with a recent\cite{TK} Monte--Carlo result, $\am=-0.2415(19)$,
obtained using VEGAS, in preference to RIWIAD (which gave\cite{KNO}
a grossly discrepant value, amended\cite{TK} in the light of a
renormalization--group analysis\cite{BKT}). In visual terms, the comparison
is
\vspace{-7mm}
\setlength{\unitlength}{0.0125cm}

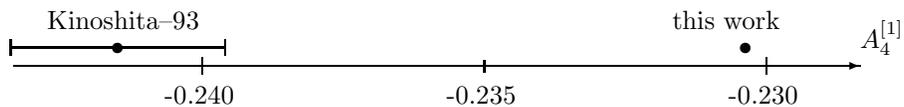
\begin{picture}(900,200)
\put(0,50){\vector(1,0){900}}
\put(200,41){\line(0,1){18}}
\put(500,44){\line(0,1){12}}
\put(800,41){\line(0,1){18}}
\put(160,10){-0.240}
\put(460,10){-0.235}
\put(760,10){-0.230}
\put(900,70){$\am$}
\put(-4,70){\line(1,0){228}}
\put(-4,62){\line(0,1){16}}
\put(224,62){\line(0,1){16}}
\put(110,70){\circle*{10}}
\put(35,90){Kinoshita--93}
\put(778,70){\circle*{10}}
\put(700,90){this work}
\end{picture}
\newpage

To verify that this discrepancy is not an artifact of
the Pad\'e method, we also tried
a hypergeometric\cite{GG2,BIS} method,
i.e.\ a polynomial fit to
$\tilde\rho_3^{[1]}(t)/\rho_3^{[{\rm c}]}(t)-\frac{1}{24}\pi^4
=\sum_{k>1} c_k\,v^k$,
with 5 coefficients, fixed by
$R(n)-\frac{1}{24}\pi^4=\sum_k c_k\left(\frac12\right)_{n+2}/
\left(\frac{k+1}{2}\right)_{n+2}$.
As might be expected, the resultant fit
to the spectral density $\rho_3^{[1]}(t)$
was less smooth than in our Pad\'e methods.
Nevertheless, the shift in the value for $\am$
was two orders of magnitude less than the disagreement
with the Monte--Carlo result.

In conclusion, we stress that the analytical data of Eqs~(1--6)
exhibit a high degree of internal consistency, making it most unlikely that
any of them is in error. Pad\'e approximants
for the mapping~(\ref{M}) of the
well--behaved function~(\ref{P}) enable us to evaluate
the muon--anomaly contribution~(\ref{I}) with an uncertainty of
2~parts in $10^5$.
Our result is in reasonable agreement with a recent, lower--precision,
Monte--Carlo re--evaluation\cite{TK},
whose uncertainties appear to have
been underestimated by a factor of 6, which
is a great improvement on the situation
revealed by a previous discrepancy between analytical\cite{BKT}
and numerical\cite{KNO} work.

\vspace*{-3pt}
\nonumsection{Acknowledgements}
\vspace*{-0.5pt}\noindent
We are grateful to K.\ G.\ Chetyrkin, A.\ L.\ Kataev and V.\ A.\ Smirnov,
for discussions of asymptotic behaviour,
to M.\ B.\ Voloshin,
for discussion of threshold behaviour,
to T.\ Kinoshita,
for correspondence on the muon anomaly,
and to A.\ C.\ Hearn,
for adapting REDUCE\cite{RED} to suit the needs of RECURSOR.
(The file to update REDUCE~3.5 is obtained by asking
reduce-netlib@rand.org to `send patches.red'.)

\nonumsection{References}

\end{document}